\documentclass[10pt,a4paper,twoside]{article}
\usepackage{epsfig}

\begin{document}

{\LARGE \bf Normalization of polarized spectral contours for high-accuracy magnetic field measurements}

\bigskip

S.~Plachinda$^1$, D.~Shulyak$^2$, N.~Pankov$^1$

$^1$Crimean Astrophysical Observatory of RAS, Nauchny, Russia, psi@craocrimea.ru
$^2$Institute for Astrophysics, Georg-August University, Friedrich-Hund-Platz 1, D-37077 G$\ddot{o}$ttingen, Germany

\bigskip

{\bf Abstract.} 

Designed at the Crimean Astrophysical Observatory the SL-method of stellar magnetic field measurement is based on the calculation of the distance between the centers of gravity of the left and right circularly polarized components of the spectral line. The SL-method is free from the limitations of the Least Squares Deconvolution (LSD) method, namely its requirement for weak magnetic fields, weak spectral lines, the same shape of polarized profiles. Also, this method is free from model limitations of the Principal Component Analysis (PCA), Nonlinear Deconvolution with Deblending (NDD) and Zeeman Component Decomposition (ZCD) also.
The SL-method  involves the calculation of the magnetic field using individual spectral lines and the subsequent formation of uniform arrays of the magnetic field measurements. The formation of the arrays is based on different criteria, e.g. using spectral lines of a chemical element, assumption of the same physical conditions in the areas of the spectral line formation, a statistically significant difference between the values of the magnetic field calculated from different arrays of spectral lines etc.
If there are several arrays of spectral lines, it is necessary to obtain normalized polarized profiles using predetermined parameters for each array: the magnitude of the field, wavelength for normalization, Lande factor and the central residual intensity of the line. 
In this paper, we present a method of normalization of polarized profiles based on the dependence of the residual intensity of the observed polarized profiles on their central residual intensity (COntour Algorithm of the Line Approximation: COALA). The SL and COALA methods strong depend on the signal-to-noise ratio and a number of used unblended spectral lines.
The resulting normalized profiles and the Stokes V can be used as the initial data in the study of the global and local topology of the magnetic field of a star. In this work the application of the proposed method is demonstrated for synthetic polarized spectra and for circularly polarized spectra of the yellow subgiant $\beta$ Aql (Sp G8 IV) obtained on 2014 October 06.

\bigskip


\bigskip

{\bf 1. Introduction}

\medskip

High-resolution spectropolarimetry allows us to measure weak magnetic fields with strength up to 1 G. The typical set of tools for such measurements include medium or large diameter telescope, echelle spectrograph, spectropolarimeter (Stokesmeter) with input rotating quarter-wave or half-wave plate, CCD detector, and software for magnetic field calculation (see, for example, Semel et al. 1993, Donati et al. 1997, Plachinda \& Tarasova 1999, Plachinda 2004, Butkovskaya \& Plachinda 2007, Plachinda 2014). The Stokesmeter allows us to measure all four Stokes parameters  (see in details Egidio Landi Degl’innocenti and Marco Landolfi  2005). Knowing the Stokes parameters at different phases of the rotation period, it is possible to reconstruct the topology of the magnetic field of a star using model calculations (see, for example Donati 2001, Kochukhov \& Piskunov 2002).

Most often, using spectropolarimetry, circular polarization in the profiles of spectral lines is investigated (Stokes $I$ and $V$). The displacement of circularly polarized components relative to the central wavelength of the spectral line provides information about the longitudinal component of the magnetic field of the star. This is the simplest way to study in the first approximation the geometry of the global magnetic field of a star.

The integral flux of a radiation from the whole visible hemisphere of a star is recorded in the spectropolarimetric observations. Therefore, unlike solar observations, the surface inhomogeneities of physical conditions and the parameters of local magnetic fields  remain unknown. As a rule, the global magnetic field of a star is described by a dipole or multipole. Sometimes such global field is called uniform, meaning that the magnetic force lines crossing a stellar surface element are parallel and the field vector does not change with respect to the line of sight during exposure.

Local magnetic fields are the fields of any spatial scales that can be distinguished against the background of the global magnetic field. Local magnetic fields on the Sun are observed at various spatial scales. They are found to be variable with different time scales from some minutes up to the twenty-two-year Hale's cycle of magnetic activity. 

Unlike the Sun, large-scale and small-scale magnetic fields of stars with convective shells are not yet studied in enough detail. In general, the atmospheres of different types of single, non-degenerated stars in the "quiet stage" are characterized by a complex of physical conditions:

\begin{enumerate}
	\item spatial inhomogeneity of temperature;
  \item spatial inhomogeneity of gravity;
  \item temperature change with depth;
  \item microturbulence change with depth;
  \item density and velocity change with depth;
  \item spatial chemical inhomogeneity;
  \item stratification of chemical elements with depth;
  \item inhomogeneity of surface magnetic field.
\end{enumerate}

In addition, each star has its own additional features according to its evolutionary status and non-stationary processes in the atmosphere. Other indicators of complex physical processes in the stellar atmosphere are different asymmetry and different central residual intensity of the circularly polarized profiles of the same spectral line (D.N. Rachkovsky, private communication). Thus, for high-precision spectropolarimetric measurements of stellar magnetic fields, preliminary calculation of the magnetic field by individual spectral lines is highly desirable.

Using blends along with unblended spectral lines to form a “mask” for the magnetic field calculation (LSD by Donati et al. 1997, PCA by Semel et al. 2006, NDD by Sennhauser et al. 2009,  ZCD by Sennhauser \& Berdyugina 2010, Kochukhov et al. 2010, Tkachenko et al. 2013) allow us to establish effectively the parameters of the global magnetic field of a star due to the high signal-to-noise ratio. To use these multiline methods for correct magnetic field calculation, it is necessary to involve additional independent selection criteria for spectral lines and to create an individual  “mask” of each subarray. Unlike these multiline methods, which use different initial model approximations, the 
SL-method works with individual observed lines (Plachinda \& Tarasova 1999, Plachinda 2004, Butkovskaya \& Plachinda 2007, Plachinda 2014), so it is free from any restrictions. But the SL-method needs a higher signal-to-noise ratio in the spectra because it does not work, as a rule, with blends of lines.

In this paper, we briefly describe the technique of normalization of the polarized profiles in observed spectra after processing by the SL-method. The main goal of this method of normalization is to calculate the  normalized polarized profile using the selected subset of observed spectral lines. For an illustration of the normalization the circularly polarized profiles in spectra of $\beta$ Aql (Sp G8 IV) obtained on 2014 October 06 at CrAO are used.

\bigskip

{\bf 2. Algorithm COALA}

\medskip

\begin{figure}[ht]
\begin{center}
\includegraphics*[height=8cm]{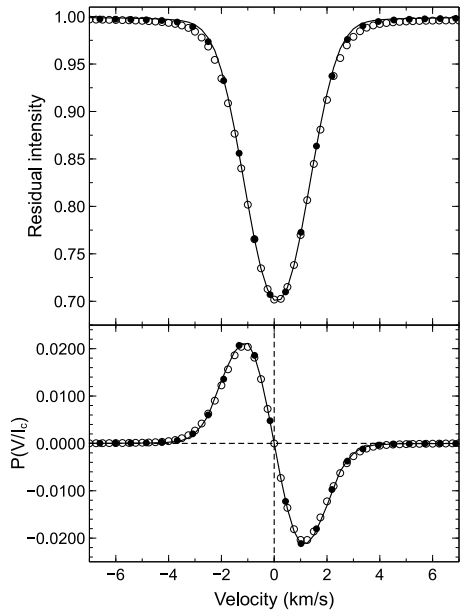}
\caption{
Mean, origin and normalized line profiles of the same circular polarization (top panel) and circular  polarization (bottom panel). The line contour averaged over 19 lines in the same reference system is shown by open circles. Closed circles show the initial contour of the line 5116.0427 \AA. Solid curve shows normalized contour: 527 spectral lines were chosen to solve the system of functional dependencies (1),  Lande factor is equal to 1.5, wavelength is equal to 5116.0427~\AA, central residual intensity for normalization is equal to 0.700. Each point of this contour was calculated using system of functional dependencies (1).}
\end{center}
\end{figure}

In order to obtain normalized polarized contours, it is necessary to combine the observed line profiles into one reference frame using several required parameters: the Lande factor, the wavelength and the magnetic field. According to the accepted observation technique, from two exposures with the oppositely oriented quarter-wave plate and “mask” of spectral lines four normalized polarized profiles and, correspondingly, the Stokes $V$ can be obtained. The control test of the correctness of the obtained result is the coincidence of the value of the initial magnetic field and the value of the magnetic field calculated from the obtained normalized contours.

The algorithm of calculations can be represented as follows.

1. The subset of the total array of magnetic field values obtained for each spectral line (the SL-method) is analyzed by comparing the statistical errors and the errors determined by the Monte Carlo method. If the difference between these two types of errors is statistically significant, then the distribution of the magnetic field values is analyzed. In addition, if the signal-to-noise ratio is sufficient, the Stokes $V$ profiles of all lines are analyzed also. According to the results of the analysis, subsets of the magnetic field values are formed, which are statistically significantly different from each other by their mean and form of Stokes $V$.

2. In the selected subset of spectral lines, the asymmetry of the polarized contours is checked separately for each polarization using bisectors. Line profiles, the asymmetry of which exceeds the permissible level, are excluded from consideration.

3. All spectral lines are transferred to the selected reference frame, which is determined by the average magnetic field of this subset of lines, selected Lande factor and selected wavelength.

4. Separately for all contours of each polarization of two types of exposures we calculate
	
\begin{equation}
r_{n,k} = f(R^{center}_{k}),
\end{equation}

where $r_{n,k}$ - residual intensity for each $n$-th point of the $k$-th contour and $R^{center}_{k}$ - residual intensity at the wavelength of the center of gravity of the $k$-th contour.

Because the elimination of the blended and unusual profiles is desirable, we use least squares polynomial fitting to seek for a solution to dependency (1) in the form of the limit

$$r_{n,k} = \lim_{f(R)\to max} f(R^{center}_{k}).$$

In the frame of our testing experience, the best fitting solution to dependence (1) is the third or fourth degree polynomial function.

5. According to the obtained dependencies (1) for a given central residual intensity, the normalized contours in each polarization for both exposures are found.

6. Using the normalized contours, the Stokes $V$ and the longitudinal component of the magnetic field are found.

This method is called COALA (COntour Algorithm of the Line Approximation). Normalization errors of method COALA are strongly demanding to the signal-to-noise ratio and to the number of spectral lines of different depths. In practice, it sometimes allowed us to use spectral lines if the blending does not exceed 10-15\% of the equivalent line width.

To illustrate the application of the COALA algorithm (see Figure 1, solid curve), the synthetic spectrum (Khan \& Shulyak 2006) was calculated for the center of the visible surface of a star in the range from 5000 to 6300 \AA. The radial component of the magnetic field was set to 100 G, and the azimuthal and meridional components were set to zero. Model atmospheres (Shulyak et al. 2004) close to the solar one was calculated for $T_{\mathrm{eff}}$ = 5777 K, lg $g$ = 4.4, $\xi$ = 0.7 km/s.

To solve the system of functional dependencies (1), 527 spectral lines were chosen within the central residual intensities 0.117 - 0.915. Normalization parameters were set as follows: Lande factor for normalization is equal to 1.5, central residual intensity for normalization is equal to 0.700, velocity step for normalization is equal to 0.25, the wavelength for normalization is equal to 5116.0427~\AA, minimal central residual intensity is equal to 0.69, maximal central residual intensity is equal to 0.71. In total, 19 lines from the source list got into these frames, including the line 5116.0427~\AA$\;$whose Lande factor $Z_{\mathrm{e}}$ = 1.5 and central residual intensity $\cong$0.70.

The generation of noise in synthetic spectra was not performed. Therefore, deviations of the final result from the initial conditions are demonstrated by errors of the method caused by the sequence of iterations with approximations by $n$-degree parabolas and third-order splines. As a rule, the errors of the method do not exceed $\sim$2\%. We can see quite well the coincidence in the case of weak lines.

In contrast to weak lines, the contours of deep saturated lines (optically thick lines) have different gradients in the wings. In this case, to obtain the average contour it is necessary to select contours of the same shape, otherwise the errors in determining the mean contour will be artificially increased.

\bigskip

{\bf 3. $\beta$ Aql }

\medskip
The application of the COALA is demonstrated using the spectropolarimetric observations of the yellow subgiant $\beta$ Aql. Butkovskaya et al. (2017) have detected the 969-day activity cycle and have studied the behavior of the magnetic field of this star during the activity cycle. The authors also found the rotation period of the star, $P_{\mathrm{rot}}$ = 5.08697 $\pm$ 0.00031 days, and reported rotation modulation of the global magnetic field with amplitude 3.3 $\pm$ 0.5 G. Long-term observations allowed them to register in some nights the strengthening of the magnetic field of the star. The nature of the strengthening is still questionable.

\begin{figure}[ht]
\begin{center}
\includegraphics*[height=8cm]{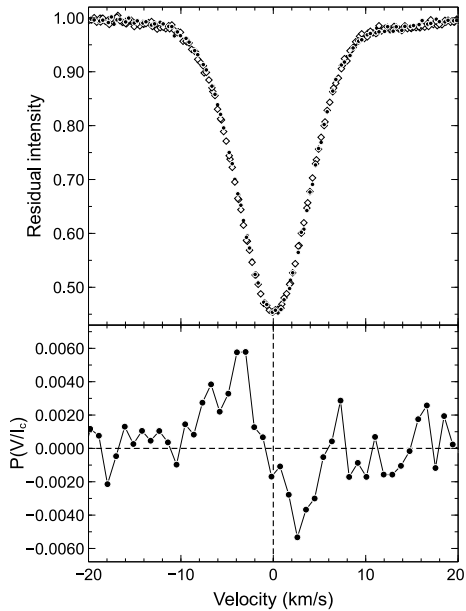}
\caption{
Top: $\beta$ Aql spectrum in the region of Fe I 6219.2810 \AA. Different circularly polarized line profiles are shown by different symbols (2 exposures, 4 profiles); the input quarter-wave plate was rotated by 90º before the second exposure. Bottom: the polarization distribution.}
\end{center}
\end{figure}

We have analyzed spectropolarimetric observations of $\beta$ Aql obtained on 2014 October 06 (rotation phase $\phi_{\mathrm{rot}}$ = 0.489), when the maximal longitudinal magnetic field, $B$ = 23.6 $\pm$ 1 G, was registered. The global magnetic field on this date was expected to be equal to $B_{\mathrm{e}}$ $\sim$ $-$4 G. The standard errors and the errors determined by the Monte Carlo method differ with a certainty of 100\%. This means that the entire array of single measurements of the magnetic field is inhomogeneous. We found, one subset of spectral lines exhibits a large magnetic field up to some tens G ("strong field") whereas other lines demonstrate weak statistically insignificant field ("weak field").

Then the total array of spectral lines was divided into two subarrays. The first array collected spectral lines whose magnetic fields deviate more than 3$\sigma$ from the rotational magnetic curve. The second subarray collected spectral lines whose magnetic fields did not show significant deviations from the rotational magnetic curve.

\begin{figure}[ht]
\begin{center}
\includegraphics*[height=8cm]{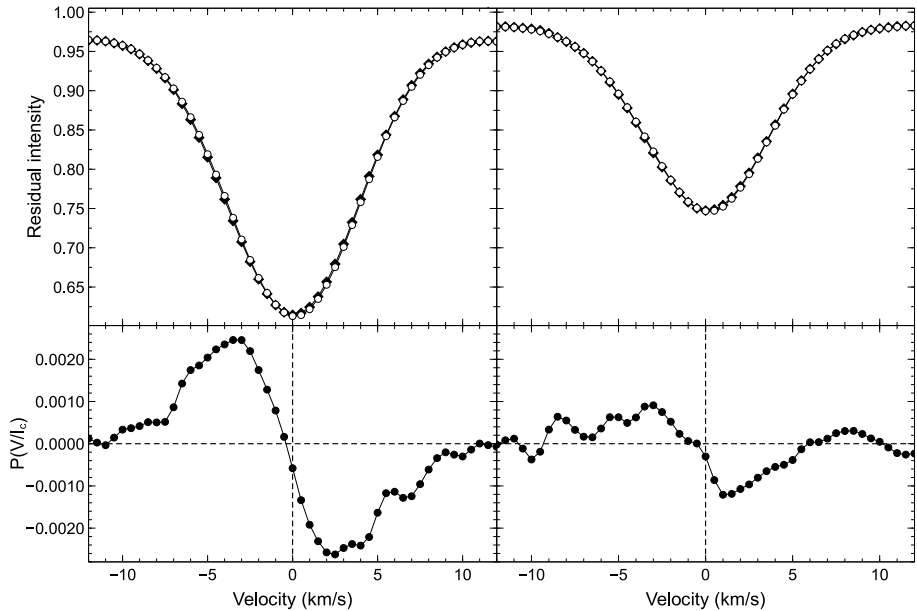}
\caption{
Left side of the picture: ''strong field''; right side of the picture: ''weak field''.}
\end{center}
\end{figure}

Normalized to the continuum origin spectrum around line Fe I 6219.2810 \AA$\;$($Z_{\mathrm{e}}$ = 1.6600, $R_{\mathrm{c}}$ = 0.450) is shown on the upper panel of Figure 2 as an example of a ''strong field''. Polarization distribution for the same spectral region is shown on the bottom panel of Figure 2. The signal-to-noise ratio in the continuum is $\sim$280$-$333 for different exposures. The field measured from this line $B_{\mathrm{e}}$ = 20.1 $\pm$ 3.4 G (the error is calculated by the Monte Carlo method).

The mean of the magnetic field, normalized profiles and normalized polarization for both subarrays of spectral lines were calculated. A total of 64 lines were used in the case of a "strong field" ($<B_{\mathrm{e}}>$ = 38.8 $\pm$ 1.8 G) and 39 lines in the case of a "weak field" ($<B_{\mathrm{e}}>$ = 15.0 $\pm$ 2.4 G). On this night, three measurements of the magnetic field were made for each line, which increased the accuracy of the normalized line profiles and their circular polarization.

On the left side of Figure 3, the upper panel shows profiles averaged over all lines and normalized to the wavelength 5500.0000 \AA$\;$and $Z_{\mathrm{e}}$ = 1.000 for the "strong field" case. The central residual intensity of the mean polarized profiles is $R_{\mathrm{c}}$ $\sim$ 0.614. The bottom panel shows the circular polarization. Errors have not been shown since the difference in the depths of the original line profiles is considerable. The extrema of polarization are close in absolute value: $P_{\mathrm{max}}$ $\sim$ 0.0024, $P_{\mathrm{min}}$ $\sim$ $-$0.0026.

On the right side of Figure 3, upper panel shows line profiles averaged over all lines and normalized to the wavelength 5500.0000 \AA$\;$and $Z_{\mathrm{e}}$ = 1.000 for the "weak field" case. The central residual intensity of the polarized profiles is $R_{\mathrm{c}}$ $\sim$ 0.746. The bottom panel shows the circular polarization (normalized to the intensity of the Stokes $V$). Errors have not been shown since the difference in the depths of the original profiles is considerable. The extrema of polarization are close in absolute value: $P_{\mathrm{max}}$ $\sim$ 0.0009, $P_{\mathrm{min}}$ $\sim$ $-$0.0012.

It should be noted that for both the "strong field" and the "weak field", the average polarization curves are shifted to the blue side by 0.5 km/s. Since spectral lines with a large range of central residual intensities were used in each averaging (i.e., deep and weak lines simultaneously), this shift can be a consequence of this averaging. To test this assumption, lines with residual intensity in the diapason $R_{\mathrm{c}}$ = 0.68 $-$ 0.72 were chosen. In the case of a "weak field", 5 out of 39 lines were chosen. In the case of a "strong field", 12 out of 64 lines were chosen.

The mean value of the oscillator strengths for the 12 weak absorption lines of "strong field", $<log~gf>$ = -1.215, and is much less than $<log~gf>$ of the 5 weak lines of "weak fields", $<log~gf>$ = -0.560. The average values of the field for both subarrays of the weak lines are not significantly different from the average values over the entire sample. For "strong field" $<B_{\mathrm{e}}>$ = 43.98 +/- 5.59 (38.8 +/- 1.8 G) and for "weak field" $<B_{\mathrm{e}}>$ = 11.96 +/- 7.90 (15.0 +/- 2.4 G). The mean low excitation potential for the weak lines of "strong field", $<E~low>$ = 3.7291, is practically the same as the mean low excitation potential for the weak lines of "weak field", $<E~low>$ = 3.1618.

\begin{figure}[ht]
\begin{center}
\includegraphics*[height=8cm]{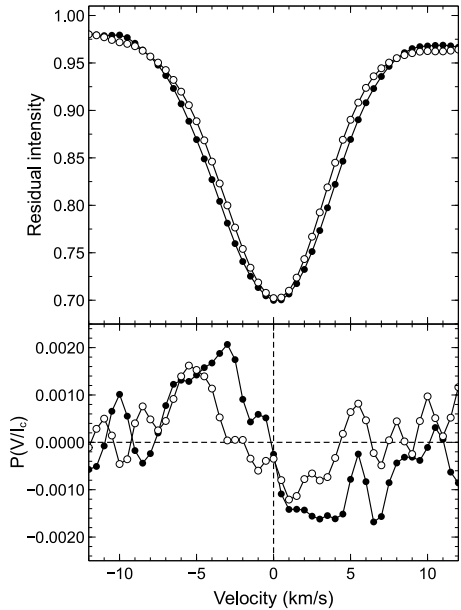}
\caption{
Comparison of line profiles (top) and polarization (bottom) for weak lines in the case of "weak field" (open circles) and "strong field" (closed circles).}
\end{center}
\end{figure}

The comparison of the average profiles for both cases is shown in Figure 4. The average contour of lines demonstrating the "strong field" is broadened in comparison with the average line contour demonstrating the "weak field". The maximum errors of the mean $\sigma$($<r_i>$) $\sim$ 0.08 at the edges of the spectral range are caused by the blending of the profile by neighboring lines. The errors inside the profile are less than the size of the symbols and fall in the range of 0.002 $-$ 0.003 for the "strong field" profile and 0.003 $-$ 0.004 for the "weak field" profile.

On the bottom panel, the curve marked with open circles shows the distribution of the circular polarization in the case of "weak field". The average standard error is $\sigma$ $\leq$ 0.0004. The curve marked with closed circles shows the circular polarization distribution for the "strong field"case. The average standard error is $\sigma$ $\leq$ 0.0003.

The polarization distribution in the case of "weak field" (open circles) shows an asymmetric pattern with respect to the velocity of the center of gravity of the line, whereas the polarization curve for the "strong field" case changes the sign in the zero velocity region. The shape of the polarization curves is significantly different from the classical Stokes $V$ profile observed in the case of the dipole global field configuration. It is obvious that in the considered case it would be incorrect to use the total mask of spectral lines to calculate the LSD profile.

\bigskip

{\bf Conclusions}

The main goal of the COALA method is obtaining the normalized polarized line profiles and Stokes $V$, using the sub-array of the total array of the spectral lines, which give us a homogeneous series of the magnetic field values. In the paper this method has been demonstrated using synthetic solar-like spectrum for longitudinal magnetic field of 100 G and the spectropolarimetric observations of beta Aql obtained on October 06, 2014.

When we process spectropolarimetric observations, we need the most homogeneous data for following successful modeling. The most widespread methods today are LSD and its modifications. Before using the classical LSD method we must first prepare an array of spectral lines (spectral mask). After calculation of LSD Stokes $I$ and $V$ profiles we obtain only one magnetic field value using all bulk of spectral lines possibly formed in different physical conditions. When we get magnetic field averaged over all bulk of spectral lines we can lose information about large-scale magnetic field inhomogeneities. In this case using the SL and COALA methods is preferable.

\medskip

\bigskip

{\bf Acknowledgements}
Authors are grateful to V.V. Butkovskaya for discussions and useful remarks with reference to the writing of this article.

\newpage

{\bf References}

\medskip

Butkovskaya, V., Plachinda, S., 2007, Astron. \& Astrophys., 469, 1069

Butkovskaya, V.V., Plachinda, S.I., Bondar’, N.I., Baklanova, D.N., 2017, Astron. Nachr., 338, 896

Donati, J.-F., Semel, M., Carter, B.D., Rees, D.E., Collier Cameron, A., 1997, MNRAS, 291, 658 

Donati, J.-F., 2001, Imaging the Magnetic Topologies of Cool Active Stars, in Astrotomography:
   Indirect Imaging Methods in Observational Astronomy, (Eds.) Boffin, H.M.J., Steeghs, D.,
   Cuypers, J., vol. 573 of Lecture Notes in Physics, pp. 207, Springer, Berlin; New York 

Degl’innocenti Egidio Landi, Landolfi Marco, 2005, Polarization in Spectral Lines. Published by
   Springer Netherlands

Khan, S. A., Shulyak, D. V., 2006, Astron. \& Astrophys., 448, 1153

Kochukhov, O., Piskunov, N.E., 2002, Astron. \& Astrophys., 388, 868

Kochukhov, O., Makaganiuk, V., Piskunov, N. 2010, Astron. \& Astrophys., 524, A5

Plachinda, S.I., Tarasova, T.N., 1999, Astrophys. J., 514, 402

Plachinda, S.I., 2004, NATO Science Series. Photopolarimetry in Remote Sensing. Editors: Gorden Videen, Yaroslav Yatskiv, Michael Mishchenko V. Springer, Dordrecht, 161, 351

Plachinda, S.I., 2014, Bulletin of the Crimean Astrophysical Observatory, 110, 17

Semel, M., Donati, J.-F., and Rees, D.E., 1993, Astron. \& Astrophys., 278, 231

Semel, M., Rees, D. E., Ram\'irez V\'elez, J. C., Stift, M. J., Leone, F., 2006, Solar Polarization 4, ASP Con. Ser., 358, 355 

Sennhauser, C., Berdyugina, S. V., Fluri, D. M., 2009, Astron. \& Astrophys., 507, 1711 

Sennhauser, C., Berdyugina, S. V., 2010, Astron. \& Astrophys. 522, A57 

Shulyak, D., Tsymbal, V., Ryabchikova, T., Stütz, C., Weiss, W. W., 2004, Astron. \& Astrophys.,
   428, 993

Tkachenko, A., Van Reeth, T., Tsymbal, V., Aerts. C., Kochukhov, O., Debosscher, 
J., 2013, Astron. \& Astrophys., 560, A37

\end{document}